\documentclass[aps,prb,twocolumn,showpacs]{revtex4-1}

\usepackage{graphicx}

\begin{document}

\title{First-principles study of ferroelectricity and pressure-induced\\
phase transitions in HgTiO$_3$}

\author{Alexander I. Lebedev}
\email[]{swan@scon155.phys.msu.ru}
\affiliation{Physics Department, Moscow State University, 119991 Moscow, Russia}

\date{\today}

\begin{abstract}
Ground-state structure is found and pressure-induced phase transitions up to
210~kbar are studied in mercury titanate from first principles within the
density functional theory. It is established that the $R3c$ structure
experimentally observed in HgTiO$_3$ is metastable at ambient pressure. With
increasing the hydrostatic pressure, the ground-state structure changes
following the $R{\bar 3} \to R3c \to Pbnm$ sequence. It is shown that the
appearance of ferroelectricity in HgTiO$_3$ at $P = 0$ is associated with an
unstable phonon mode. Optical and elastic properties of different phases of
mercury titanate are calculated. The quasiparticle band gap calculated in the
\emph{GW} approximation ($E_g = 2.43$~eV) agrees with experimental data better
than the value obtained in the LDA approximation (1.49~eV). Analysis of the
thermodynamic stability explains why the synthesis of mercury titanate is
possible only at high pressures.
\end{abstract}


\maketitle

\section{Introduction}

The strain engineering has already become an important technological approach
which enables to enhance the properties of many electronic materials. In
ferroelectrics, due to the large strain--polarization coupling, the strain
effects are especially important. For example, in ferroelectric thin films and
superlattices, the biaxial strain can be used to finely tune the ferroelectric,
dielectric, and piezoelectric properties of these
materials.~\cite{CurrOpinSolidStateMaterSci.9.122,AnnuRevMaterSci.37.589,
PhysSolidState.51.2324}

The influence of hydrostatic pressure on displacive phase transitions was
first analyzed by Samara \emph{et al.},~\cite{PhysRevLett.35.1767} who
explained why an increase in pressure usually decreases the temperature
of phase transitions associated with soft optical phonons at the center of
the Brillouin zone and increases the temperature of phase transitions associated
with soft phonons at the boundary of the Brillouin zone. Recently, a very
different behavior of the ferroelectric properties was discovered in PbTiO$_3$
at high pressures; the observed enhancement of the ferroelectric instability
with increasing pressure was explained by the original electronic mechanism of
ferroelectricity.~\cite{PhysRevLett.95.196804} The enhancement of
ferroelectricity at high pressures was predicted for many oxides with the
perovskite structure.~\cite{PhaseTransitions.80.385} These interesting
findings necessitate further studies of the pressure effects on the ferroelectric
properties.

Despite limited experimental data on mercury titanate, this material
exhibits interesting but contradictory ferroelectric properties. Mercury
titanate HgTiO$_3$ can be prepared from HgO and TiO$_2$ at pressures of
60--65~kbar.~\cite{JSolidStateChem.6.509,Ferroelectrics.326.117}  The
obtained crystals have a rhombohedrally distorted perovskite structure.
The observation of second harmonic generation (SHG) in HgTiO$_3$ at
300~K~\cite{JSolidStateChem.6.509} enabled to propose that mercury titanate is
non-centrosymmetric and its space group is $R3c$. However, because of limited
accuracy, the atomic coordinates in Ref.~\onlinecite{JSolidStateChem.6.509}
were determined only for centrosymmetric structure $R{\bar 3}c$.
Subsequent studies of dielectric properties of mercury
titanate~\cite{Ferroelectrics.326.117,Ferroelectrics.337.71} did not found
sharp dielectric anomalies: a strongly asymmetric broad peak with a maximum
dielectric constant of $\sim$800 at about 220~K and a noticeable hysteresis
in the heating--cooling cycle as well as weak narrow peak at about 515~K were
observed. At 300~K no dielectric hysteresis loops were observed in electric
fields up to 10$^6$~V/m.~\cite{Ferroelectrics.326.117,Ferroelectrics.337.71}
Scanning calorimetry revealed weak anomalies in the 420--480~K temperature
range,~\cite{Ferroelectrics.326.117,Ferroelectrics.337.71}  but their
temperatures differed from the temperatures of maximums in dielectric
measurements.

X-ray diffraction studies of HgTiO$_3$ under hydrostatic
pressure~\cite{Ferroelectrics.326.117,Ferroelectrics.337.71} revealed
non-monotonic behavior of the $d_{024}$ interplanar distance and of the
(104)--(110) doublet splitting at $P \approx 20$~kbar, which was explained by
a phase transition from the rhombohedral to the cubic phase. Studies of the
electronic structure~\cite{JPhysCondensMatter.22.045504} of rhombohedral and
cubic modifications of HgTiO$_3$ using the full-potential linearized augmented
plane wave (FP-LAPW) method showed that the rhombohedral
$R{\bar 3}c$ phase is a direct-gap semiconductor with the band gap energy of
$\sim$1.6~eV, whereas the cubic phase is a metal.

To resolve the inconsistency of the ferroelectric properties of HgTiO$_3$ and
predict other properties of this material, first-principles calculations
of the ground-state structure, phonon spectra, optical and elastic properties
of mercury titanate at hydrostatic pressures up to 210~kbar were performed in
this paper.

\section{Calculation details}

\begin{table*}
\caption{\label{table1}Parameters used for construction of pseudopotentials.
All parameters are in Hartree atomic units except for the $V_{\rm loc}$ energy
which is in Ry.}
\begin{ruledtabular}
\begin{tabular}{ccccccccccc}
Atom & Configuration         & $r_s$ & $r_p$ & $r_d$ & $q_s$ & $q_p$ & $q_d$ & $r_{\rm min}$ & $r_{\rm max}$ & $V_{\rm loc}$ \\
\hline
Hg   & $5d^{10} 6s^0 6p^0$   & 1.78  & 2.00  & 1.78  & 7.37  & 7.07  & 7.37 & ---  & ---  & --- \\
Ti   & $3s^2 3p^6 3d^0 4s^0$ & 1.48  & 1.72  & 1.84  & 7.07  & 7.07  & 7.07 & 0.01 & 1.41 & 2.65 \\
O    & $2s^2 2p^4 3d^0$      & 1.40  & 1.55  & 1.40  & 7.07  & 7.57  & 7.07 & ---  & ---  & --- \\
\end{tabular}
\end{ruledtabular}
\end{table*}

The calculations were performed within the first-principles density-functional
theory (DFT) using pseudopotentials and a plane-wave expansion of wave functions,
as implemented in \texttt{ABINIT} software.~\cite{abinit3}   The local density
approximation (LDA) for the exchange-correlation functional was used. Optimized
separable nonlocal pseudopotentials~\cite{PhysRevB.41.1227}
were constructed using the \texttt{OPIUM} program.~\cite{OPIUM}  The parameters
used for the construction of pseudopotentials are given in Table~\ref{table1}
(for meaning of the parameters see Ref.~\onlinecite{PhysRevB.41.1227}); for Ti
atom the local potential correction~\cite{PhysRevB.59.12471} was added.

The pseudopotentials for Ti and O atoms were non-relativistic (they have been
already tested and used in Ref.~\onlinecite{PhysSolidState.51.362}), the
pseudopotential for Hg was constructed using scalar-relativistic generation
scheme. The plane-wave energy cut-off used in the calculations was 30~Ha
(816~eV). Integration over the Brillouin zone was performed on
8$\times$8$\times$8 Monkhorst--Pack mesh for the dielectric phases and
12$\times$12$\times$12 mesh for the metallic phases with one formula unit
in the unit cell. For larger unit cells, equivalent $\mathbf{k}$-point density
was used. The total energy was converged to less than 10$^{-10}$~Ha. The
structure relaxation was stopped when the Hellmann--Feynman forces were
below $5 \times 10^{-6}$~Ha/Bohr (0.25~meV/{\AA}).

The convergence tests with a number of irreducible $\mathbf{k}$-points
increased by 2--4~times have shown that the changes in the calculated total
energies did not exceed 0.05~meV for the dielectric phases and 0.35~meV
for the metallic phases. The changes in the phonon frequencies were less
than 0.01~cm$^{-1}$ for hard modes and less than 1~cm$^{-1}$ for soft modes.
An increase in the energy cut-off also had a little effect on the calculated
data: the change in the relative energies of different phases was less than
0.15~meV when the cut-off energy was increased to 35~Ha.

To test the quality of pseudopotential for Hg, the calculations for orthorhombic
and rhombohedral polymorphs of HgO were performed. Among them the orthorhombic
modification (the montroydite mineral) had the lowest total energy. The
calculated lattice parameters of these phases
($a = 3.4663$~{\AA}, $b = 6.6253$~{\AA}, $c = 5.3013$~{\AA} for the orthorhombic
phase and $a = 3.5092$~{\AA}, $c = 8.5417$~{\AA} for the rhombohedral phase) were
in reasonable agreement with
experiment~\cite{SpringerMaterials} ($a = 3.5215$~{\AA}, $b = 6.6074$~{\AA},
$c = 5.5254$~{\AA}; $a = 3.577$~{\AA}, $c = 8.681$~{\AA}). Phonon spectra were
calculated with the same interpolation scheme as was used for other titanates
of group~II elements in the periodic table.~\cite{PhysSolidState.51.362}

The quasiparticle band gap in HgTiO$_3$ was calculated using the so-called one-shot
\emph{GW} approximation.~\cite{RevModPhys.74.601,PhysStatSolidiB.246.1877}
The Kohn--Sham wave functions and energies calculated within DFT-LDA were used
as a zeroth-order approximation. The dielectric matrix
$\epsilon_{\mathbf{GG'}}(\mathbf{q},\omega)$ was computed for
4$\times$4$\times$4 $\mathbf{q}$-mesh from the independent-particle
polarizability matrix $P^0_{\mathbf{GG'}}(\mathbf{q},\omega)$ calculated for
4285~reciprocal-lattice vectors $\mathbf{G}(\mathbf{G'})$, 42~occupied and
158~unoccupied bands. The contribution of higher-lying bands was taken into
account using the approach proposed in Ref.~\onlinecite{PhysRevB.78.085125}.
The dynamic screening was described using the Godby--Needs plasmon-pole model.
The components of wave functions with kinetic energy below 24~Ha were used
in these calculations. The energy correction to the DFT-LDA solution was
computed as diagonal matrix elements of $\Sigma - E_{xc}$ operator, where
$\Sigma = GW$ is the self-energy operator, $E_{xc}$ is the exchange-correlation
energy operator, $G$ is the Green's function, and $W = \epsilon^{-1}v$ is the
screened Coulomb interaction. In the calculations of $\Sigma$, the components
of wave functions with kinetic energy below 24~Ha for both exchange and
correlation parts of $\Sigma$ were used.

\section{Results}

\subsection{Ground-state structure at $P = 0$}

\begin{figure}
\centering
\includegraphics{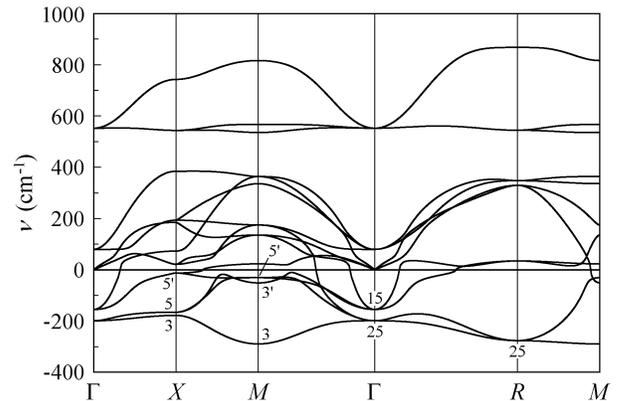}
\caption{Phonon dispersion curves for HgTiO$_3$ in the cubic $Pm3m$ phase. The
labels near the curves denote the symmetry of unstable modes. The absence
of LO--TO-splitting at the $\Gamma$ point is caused by metallic character
of this phase.}
\label{fig1}
\end{figure}

The calculated phonon spectrum of HgTiO$_3$ in the perovskite cubic phase
(space group $Pm3m$) is shown in Fig.~\ref{fig1}. It is seen that two types
of instability appear simultaneously in this phase: the stronger one
associated with the deformation and rotation of the oxygen octahedra
(the $\Gamma_{25}$--$X_3$--$M_3$--$\Gamma_{25}$--$R_{25}$--$M_3$ branch)
and the weaker one associated with the ferroelectric (antiferroelectric at
the boundary of the Brillouin zone) instability (the
$\Gamma_{15}$--$X^{\prime}_5$--$M^{\prime}_3$--$\Gamma_{15}$ branch). The
absence of LO--TO-splitting at the $\Gamma$ point is caused by the metallic
band structure of cubic HgTiO$_3$.

\begin{table}
\caption{\label{table2}Energies and volumes per one formula unit for different
distorted phases of HgTiO$_3$ at $P = 0$. The energy of the cubic phase
is taken as the energy reference. The phase with a minimum specific energy
and minimum specific volume are denoted by bold values.}
\begin{ruledtabular}
\begin{tabular}{cccc}
Unstable   & Space         & Energy,   & Volume, \\
mode       & group         & meV       & {\AA}$^3$ \\
\hline
---           & $Pm3m$        & 0       & 57.573 \\
$X_3$         & $P4_2/mmc$    & $-$88   & 57.275 \\
$\Gamma_{15}$ & $R3m$         & $-$94   & 58.923 \\
$\Gamma_{15}$ & $P4mm$        & $-$122  & 59.444 \\
$\Gamma_{25}$ & $P{\bar 4}m2$ & $-$139  & 57.088 \\
$\Gamma_{15}$,
$\Gamma_{25}$ & $Amm2$        & $-$151  & 59.749 \\
$X_5$         & $Pmma$        & $-$202  & 57.916 \\
$X_5$         & $Cmcm$        & $-$306  & 57.867 \\
$\Gamma_{25}$ & $R32$         & $-$467  & 56.956 \\
$R_{25}$      & $I4/mcm$      & $-$778  & 56.188 \\
$M_3$         & $P4/mbm$      & $-$809  & 56.195 \\
$R_{25}+M_3$  & $Pbnm$        & $-$936  & \textbf{55.853} \\
$R_{25}$      & $Imma$        & $-$940  & 56.099 \\
$R_{25}$      & $R{\bar 3}c$  & $-$974  & 56.336 \\
\hline
$A_{2u}$      & $R3c$         & $-$982  & 56.632 \\
\hline
---           & $R{\bar 3}$ & \textbf{$-$1059} & 60.140 \\
\end{tabular}
\end{ruledtabular}
\end{table}

\begin{table*}
\caption{\label{table3}The lattice parameters and atomic coordinates in
HgTiO$_3$ phases with $R3c$, $R{\bar 3}c$, and $R{\bar 3}$ space groups
at $P = 0$, and in the $Pbnm$ phase at 141~kbar.}
\begin{ruledtabular}
\begin{tabular}{cccccccc}
\rule{0pt}{4.2mm}%
Phase   & $a$,~{\AA} & $\alpha$, deg. & Atom & Position & $x$     & $y$     & $z$ \\
\hline
$R3c$   & 5.4984     & 58.4093  & Hg   & 2a      & 0.24904 & 0.24904 & 0.24904 \\
        &            &          & Ti   & 2a      & $-$0.00333 & $-$0.00333 & $-$0.00333 \\
        &            &          & O    & 6b      & 0.66598 & $-$0.15240 & 0.25846 \\
$R{\bar 3}c$ & 5.4881 & 58.4252 & Hg   & 2a      & 0.25000 & 0.25000 & 0.25000 \\
(calc.) &            &          & Ti   & 2b      & 0.00000 & 0.00000 & 0.00000 \\
        &            &          & O    & 6e      & 0.65983 & $-$0.15983 & 0.25000 \\
$R{\bar 3}c$ & 5.4959 & 58.59   & Hg   & 2a      & 0.25 & 0.25 & 0.25 \\
(exp.)$^a$   &       &          & Ti   & 2b      & 0.0 & 0.0 & 0.0 \\
             &       &          & O    & 6e      & 0.665   & $-$0.165 & 0.25 \\
$R{\bar 3}$ & 5.8304 & 53.9320  & Hg   & 2c      & 0.36869 & 0.36869 & 0.36869 \\
        &            &          & Ti   & 2c      & 0.84974 & 0.84974 & 0.84974 \\
        &            &          & O    & 6f      & 0.55966 & $-$0.03220 & 0.19275 \\
$Pbnm$  & 5.2678 ($a$) & ---    & Hg   & 4c      & $-$0.00445 & 0.03190 & 0.25000 \\
        & 5.2983 ($b$) &        & Ti   & 4b      & 0.50000 & 0.00000 & 0.00000 \\
        & 7.5501 ($c$) &        & O    & 4c      & 0.08502 & 0.47264 & 0.25000 \\
        &            &          & O    & 4d      & 0.69594 & 0.30216 & 0.04431 \\
\end{tabular}
\end{ruledtabular}
\noindent{\footnotesize $^a$~The experimental data taken from
Ref.~\onlinecite{JSolidStateChem.6.509} were recalculated for rhombohedral setting. \hfill}
\end{table*}

To determine the structure of the ground state, the energies of different
distorted phases originating from the cubic structure were calculated for
each of the above-mentioned unstable modes taking into account their
degeneracy. As follows from Table~\ref{table2}, the $R{\bar 3}c$ phase
has the lowest energy among these phases. This phase is derived from the
$Pm3m$ structure by out-of-phase rotations of the oxygen octahedra around
three cubic axes as a result of condensation of triply degenerate $R_{25}$
mode at the boundary of the Brillouin zone ($a^-a^-a^-$ tilt system according
to Glazer's notation). The energy of this phase is even lower than that of
the $Pbnm$ phase, in contrast to other titanates of group~II elements in the
periodic table.~\cite{PhysSolidState.51.362}  It should be noted that the
overlapping of the conduction and valence bands disappears as the structural
distortions become larger, and all phases with the energy lower than $-$300~meV
are semiconductors.

\begin{table}
\caption{\label{table4}Interatomic distances in HgTiO$_3$ phases with $R3c$
and $R{\bar 3}c$ space groups.}
\begin{ruledtabular}
\begin{tabular}{ccccc}
\rule{0pt}{4.2mm}%
Pair of  & \multicolumn{3}{c}{Distance,~{\AA}}                                     & Number \\
\cline{2-4}
atoms    & \multicolumn{2}{c}{This work} & Ref.~\onlinecite{JSolidStateChem.6.509} & of bonds \\
\cline{2-3}
\rule{0pt}{4.2mm}%
         & $R3c$ & $R{\bar 3}c$ & \\
\hline
Hg--O        & 2.198        & 2.195 & 2.20(4) & 3 \\
Hg--O        & 2.698, 2.888 & 2.786 & 2.77(4) & 3+3 \\
Hg--O        & 3.172        & 3.162 & ---     & 3 \\
Ti--O        & 1.906, 2.064 & 1.977 & 1.96(4) & 3+3 \\
\end{tabular}
\end{ruledtabular}
\end{table}

The ferroelectric instability specific for the reference $Pm3m$ structure of
HgTiO$_3$ still exists in the $R{\bar 3}c$ phase. The calculations show
that two unstable modes with $A_{2u}$ and $E_u$ symmetry and frequencies of
135$i$ and 21$i$~cm$^{-1}$ are observed at the $\Gamma$ point in the phonon
spectrum of this phase. Among the corresponding ferroelectrically distorted
phases, the $R3c$ phase has the lowest energy. All phonon frequencies at
the center and at $A$, $D$, and $Z$ points on the boundary of the Brillouin
zone in this phase are positive, the matrix of elastic moduli (Sec.~\ref{sec3c})
is positive-definite, and so the $R3c$ phase is the ground-state structure.
The calculated lattice parameters and atomic coordinates in $R{\bar 3}c$ and
$R3c$ phases are given in Table~\ref{table3}. For comparison, the experimental
data for the $R{\bar 3}c$ phase~\cite{JSolidStateChem.6.509} are also included
into this table. It is seen that the calculated data for the $R{\bar 3}c$ phase
agrees well with the experimental data. The calculated interatomic distances for
the $R{\bar 3}c$ phase and calculated mean distances for the $R3c$ phase agree with
interatomic distances determined from X-ray diffraction~\cite{JSolidStateChem.6.509}
too (Table~\ref{table4}).

Along with the phases originating from the perovskite structure, other possible
structures, in particular, the ilmenite one characteristic for titanates of
group~II elements---MgTiO$_3$, ZnTiO$_3$, and CdTiO$_3$~\cite{SpringerMaterials}%
---should be tested. The calculation showed that at ambient pressure ($P = 0$)
the ilmenite structure of HgTiO$_3$ with $R{\bar 3}$
space group has the lowest energy among all considered phases (Table~\ref{table2}).
The fact that $R3c$ or $R{\bar 3}c$ structures are observed in X-ray diffraction
enables to suppose that these phases are metastable. This metastability is
evidently associated with large difference between $R3c$ ($R{\bar 3}c$) and
$R{\bar 3}$ structures, both in the lattice parameter and in the rhombohedral
angle (Table~\ref{table3}). The phase transition between so different structures
should be of the first order, for which a wide region of metastability is
characteristic. The reason why all investigated samples had the metastable $R3c$
($R{\bar 3}c$) structure is that the samples have been prepared at 60--65~kbar,
at which the $R{\bar 3}c$ phase is thermodynamically stable (Sec.~\ref{sec3d}).
The energies of two more possible hexagonal phases, one with the 2H BaNiO$_3$
structure and other with 6H BaTiO$_3$ structure (both have $P6_3/mmc$ space
group), are, respectively, by 269~meV and 73~meV higher than the energy of the
cubic $Pm3m$ phase.

\subsection{Origin of ferroelectric instability}

We now discuss the ferroelectric properties and nature of the ferroelectric phase
transition in HgTiO$_3$. As the changes in the Hg--O bond lengths accompanying
the ferroelectric phase transition do not exceed 0.1~{\AA} and the difference
in energy of the $R3c$ and $R{\bar 3}c$ phases is only 8.1~meV, the Curie
temperature in HgTiO$_3$ cannot be too high and will not exceed 300~K. Therefore
we can associate the Curie temperature with the temperature of the first maximum
in the dielectric constant, which
was observed at 220~K.~\cite{Ferroelectrics.326.117,Ferroelectrics.337.71}
This interpretation is confirmed by the absence of dielectric hysteresis loops
at 300~K. The fact that SHG signal was observed in Ref.~\onlinecite{JSolidStateChem.6.509}
at 300~K can be explained by the presence of defects; the cause of the defect
formation in HgTiO$_3$ will be discussed in Sec.~\ref{sec3e}.

Calculation of the dynamical matrix for the $R{\bar 3}c$ phase shows that the
$zz$-element of the on-site force constant for Hg atoms is small but positive
(0.0263~Ha/Bohr$^2$), and so these atoms cannot be considered as off-center ions.
The analysis of the eigenvector of the ferroelectric $A_{2u}$ mode in this
phase shows that the displacements of Hg atoms in this mode is 22~times smaller
than that of Ti atoms. This means that collective displacements of titanium
atoms against oxygen atoms are responsible for the ferroelectric instability.
Weak ferroelectric activity of Hg atoms is confirmed by small values of their
Born effective charges: $Z^*_{xx} = Z^*_{yy} = 3.20$, $Z^*_{zz} = 2.42$; they
slightly exceed the nominal charge of the ion. For Ti atoms the corresponding
values are $Z^*_{xx} = Z^*_{yy} = 7.85$, $Z^*_{zz} = 7.92$.

The calculated static dielectric constant at 0~K in the $R3c$ phase is
almost isotropic ($\varepsilon_{xx} = 97$, $\varepsilon_{zz} = 101$).%
    \footnote{In the
    $R{\bar 3}$ phase, the dielectric constant is considerably lower:
    $\varepsilon_{xx} = 28$, $\varepsilon_{zz} = 27$.}
Its value is compatible with a maximum dielectric constant of $\sim$800 observed
in the experiment at 220~K.~\cite{Ferroelectrics.326.117,Ferroelectrics.337.71}
The calculated spontaneous polarization in the $R3c$ phase turns out unexpectedly
large, $P_s = 0.37$~C/m$^2$. Apparently, this is a result of large effective
charge of the $A_{2u}$ mode in the paraelectric $R{\bar 3}c$ phase
($Z^*_{\rm eff} = {}$12.66).

\subsection{\label{sec3c}Optical and elastic properties of HgTiO$_3$}

The electronic structure calculations performed in this work within the
DFT-LDA approach confirm earlier result~\cite{JPhysCondensMatter.22.045504}
obtained in the generalized gradient approximation (GGA) that the $R{\bar 3}c$
phase is a direct-gap
semiconductor with the band gap of 1.6~eV and the band extrema located at the
$\Gamma$ point. In our calculations, the band gap of HgTiO$_3$ at $P = 0$ is
$E_g^{\rm LDA} = 1.49$~eV and its pressure coefficient is
$dE_g^{\rm LDA}/dP = +0.44$~meV/kbar. However, both these results disagree with
the experimental fact that HgTiO$_3$ crystals are of light yellow
color.~\cite{JSolidStateChem.6.509}

It is well known that the DFT always underestimates the energy band gap. The
\emph{GW} approximation~\cite{RevModPhys.74.601,PhysStatSolidiB.246.1877}
based on the many-body perturbation theory is an approach that enables to
obtain $E_g$ values in good agreement with the experiment. The calculations
carried out in this work in this approximation gave the band gap energy
$E_g^{GW} \approx 2.43$~eV for the $R{\bar 3}c$ phase of HgTiO$_3$. This
value agrees with the color of the samples much better than the $E_g$ values
calculated in the LDA (1.49~eV, this work) and GGA (1.6~eV,
Ref.~\onlinecite{JPhysCondensMatter.22.045504}) approximations. The optical
dielectric constant of HgTiO$_3$ which takes into account the local field
effects is $\epsilon_\infty = 9.29$.

\begin{table*}
\caption{\label{table5}Calculated frequencies of optical phonons at the $\Gamma$
point of the Brillouin zone for HgTiO$_3$ phases with $R3c$, $R{\bar 3}$, and
$Pbnm$ space groups (the latter---at $P = 147$~kbar).}
\begin{ruledtabular}
\begin{tabular}{cccccccccc}
\multicolumn{4}{c}{$R3c$ structure} & \multicolumn{6}{c}{$Pbnm$ structure} \\
\hline
Mode    & $\nu$, cm$^{-1}$ & Mode & $\nu$, cm$^{-1}$ & Mode & $\nu$, cm$^{-1}$ & Mode & $\nu$, cm$^{-1}$ & Mode & $\nu$, cm$^{-1}$\\
\hline
$A_1$   &  78 & $E$     &  81 & $A_g$    &  68 & $B_{1g}$ & 439 & $B_{1u}$ & 530 \\
        & 181 &         & 121 &          & 113 &          & 502 & $B_{2u}$ &  38 \\
        & 379 &         & 139 &          & 144 &          & 782 &          &  86 \\
        & 476 &         & 165 &          & 277 & $B_{2g}$ & 104 &          & 141 \\
$A_2$   &  62 &         & 274 &          & 417 &          & 266 &          & 190 \\
        & 347 &         & 312 &          & 462 &          & 450 &          & 345 \\
        & 355 &         & 443 &          & 559 &          & 542 &          & 356 \\
        & 417 &         & 495 & $A_u$    &  65 &          & 819 &          & 431 \\
        & 753 &         & 515 &          &  74 & $B_{3g}$ & 118 &          & 496 \\
\cline{1-4}
\multicolumn{4}{c}{$R{\bar 3}$ structure} && 108 &       & 226 &          & 524 \\
\cline{1-4}
$A_g$ &  73 & $E_g$ &  94     &          & 141 &          & 353 & $B_{3u}$ &  58 \\
      & 204 &       & 189     &          & 303 &          & 539 &          & 114 \\
      & 306 &       & 311     &          & 375 &          & 732 &          & 165 \\
      & 440 &       & 442     &          & 498 & $B_{1u}$ &  38 &          & 246 \\
      & 651 &       & 565     &          & 539 &          &  84 &          & 301 \\
$A_u$ & 133 & $E_u$ & 148     & $B_{1g}$ &  80 &          & 134 &          & 380 \\
      & 348 &       & 256     &          & 103 &          & 243 &          & 403 \\
      & 477 &       & 371     &          & 139 &          & 387 &          & 448 \\
      & 647 &       & 452     &          & 350 &          & 475 &          & 547 \\
\end{tabular}
\end{ruledtabular}
\end{table*}

To interpret experimental data of infrared (IR) and Raman studies, the phonon
frequencies calculated at the $\Gamma$ point for different phases of HgTiO$_3$
may be useful. The data for $R3c$ and $R{\bar 3}$ phases at ambient pressure
as well as for the $Pbnm$ phase at 147~kbar are presented in Table~\ref{table5}.
In the low-temperature ferroelectric $R3c$ phase, $A_1$ and $E$ modes are
active both in IR and Raman spectra. In the $R{\bar 3}$ phase, $A_u$ and $E_u$
modes are IR~active, whereas $A_g$ and $E_g$ modes are Raman active. In the
high-pressure $Pbnm$ phase (Sec.~\ref{sec3d}), the $B_{1u}$, $B_{2u}$, and
$B_{3u}$ modes are IR~active, whereas $A_g$, $B_{1g}$, $B_{2g}$, and $B_{3g}$
are Raman active.

The elastic moduli tensor in the $R{\bar 3}c$ phase is presented by seven
independent components:
$C_{11} = C_{22} = 348.0$~GPa, $C_{33} = 260.3$~GPa, $C_{12} = 178.5$~GPa,
$C_{13} = C_{23} = 149.9$~GPa, $C_{44} = C_{55} = 76.3$~GPa, $C_{66} = 84.8$~GPa,
and $C_{14} = -C_{24} = C_{56} = 18.3$~GPa. The bulk modulus calculated from these
data is $B = 205.8$~GPa, which is slightly higher than the 178~GPa value
obtained in Ref.~\onlinecite{JPhysCondensMatter.22.045504} without taking into
account the relaxation of internal degrees of freedom.

In the $R3c$ phase, seven independent components of the elastic tensor are:
$C_{11} = C_{22} = 293.9$~GPa, $C_{33} = 224.4$~GPa, $C_{12} = 159.7$~GPa,
$C_{13} = C_{23} = 117.1$~GPa, $C_{44} = C_{55} = 53.7$~GPa, $C_{66} = 67.1$~GPa,
and $C_{14} = -C_{24} = C_{56} = 0.82$~GPa. The bulk modulus in this phase is
$B = 171.3$~GPa.

\subsection{\label{sec3d}Pressure-induced phase transition}

To discuss the experimental data on the influence of hydrostatic pressure on
the structure of HgTiO$_3$,~\cite{Ferroelectrics.326.117,Ferroelectrics.337.71}
calculations of the pressure effect on the properties of these crystals were
performed. At pressure $P \ne 0$, the thermodynamically stable phase is the
phase which has the lowest enthalpy $H = E_{\rm tot} + PV$, not the lowest total
energy $E_{\rm tot}$. To compare the phases with different number of atoms
in the unit cell, we use the specific energy and the specific volume defined
per one formula unit. The calculations show that with increasing pressure, the
contribution of the $PV$ term gives about 95\% of the change in $H$ in our
crystals, and so at high pressures the phase with a minimum specific volume becomes
more stable. As follows from Table~\ref{table2}, at $P = 0$ the $Pbnm$ phase
has the minimum specific volume. Other phases, arranged in order of increasing
their specific volumes, form the following sequence: $Imma$, $I4/mcm$, $P4/mbm$,
$R{\bar 3}c$, $R3c$. The ilmenite structure $R{\bar 3}$, which has the minimum
total energy at $P = 0$, is characterized by the largest specific volume. This
enables to expect that as the pressure is increased, the ground-state structure
will change in the following sequence: $R{\bar 3} \to R3c \to Pbnm$. Moreover,
with increasing pressure, the suppression of ferroelectricity should be observed
(the $R3c \to R{\bar 3}c$ phase transition).

\begin{figure}
\centering
\includegraphics{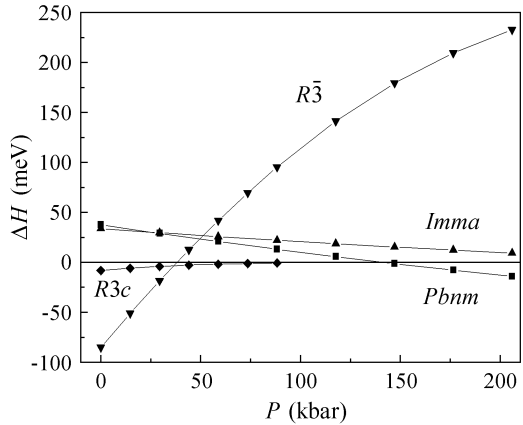}
\caption{The enthalpy differences between $R{\bar 3}$, $R3c$, $Imma$, and $Pbnm$
phases and the $R{\bar 3}c$ phase of HgTiO$_3$ as a function of hydrostatic
pressure.}
\label{fig2}
\end{figure}

The enthalpy differences between the phases under consideration and the
$R{\bar 3}c$ phase as a function of pressure are plotted in Fig.~\ref{fig2}.
It is seen that HgTiO$_3$ should undergo the $R{\bar 3} \to R3c$ phase
transition at $P = 38$~kbar and the $R3c \to Pbnm$ one at 141~kbar. As both phase
transitions are accompanied by abrupt changes in the volume of the unit cell at the
transition pressure (6\% and 0.71\%, respectively), they should be of the first
order. The $Imma$ phase, whose enthalpy at $P = 0$ is lower than that of the $Pbnm$
phase, at higher pressure becomes thermodynamically less stable and so can be
excluded from consideration. Similar behavior, when stable $R{\bar 3}$ phase
transformed under high pressure and temperature to the pressure-stabilized $Pbnm$
phase, from which the metastable $R{\bar 3}c$ phase appeared after releasing
the pressure, was observed in MnTiO$_3$,~\cite{PhysChemMinerals.16.621}
FeTiO$_3$,~\cite{PhysChemMinerals.18.244} and ZnGeO$_3$.~\cite{PhysChemMinerals.33.217}
Ferroelectric LiTaO$_3$ also exhibited the $R3c \to Pbnm$ phase transition at
high pressures.~\cite{JApplPhys.102.083503}

\begin{figure}
\begin{center}
\includegraphics{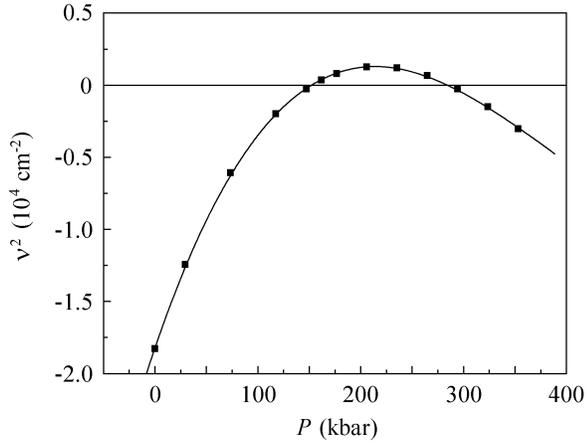}
\caption{The square of the $A_{2u}$ mode frequency in the $R{\bar 3}c$ phase
of HgTiO$_3$ as a function of hydrostatic pressure.}
\label{fig3}
\end{center}
\end{figure}

To determine the pressure of the $R3c \to R{\bar 3}c$ phase transition
accurately, the square of the $A_{2u}$ phonon frequency in the $R{\bar 3}c$
phase was plotted as a function of hydrostatic pressure (Fig.~\ref{fig3}).
It turned out that this dependence is non-monotonic and the pressure of
the phase transition is $\sim$152~kbar. This means that in the whole pressure
range before the phase transformation to the $Pbnm$ phase at 141~kbar the ground
state at $T = 0$ is ferroelectric (the $R3c$ phase). In the $Pbnm$ phase, the
ferroelectric instability is absent.

An interesting feature which is seen in Fig.~\ref{fig3} is the
reentrance of the ferroelectric instability in the $R{\bar 3}c$ phase at
pressures above 290~kbar. Similar behavior of the ferroelectric mode frequency
with increasing pressure was predicted for a number of oxides with the cubic
perovskite structure~\cite{PhysRevLett.95.196804,PhaseTransitions.80.385}
and explained by an increased
mixing of the Ti $3d$ and the O $2s$ orbitals at high pressures. The effect
observed in HgTiO$_3$ needs further investigation, but it should be noted
that it is observed in a crystal in which the oxygen environment around the Ti
atom is different from that in cubic perovskites.

The obtained results enable to propose new interpretation of the phase transition
observed in high-pressure X-ray experiments.~\cite{Ferroelectrics.326.117,
Ferroelectrics.337.71}  The specific energy of the cubic ($Pm3m$) phase, which
was considered in Refs.~\onlinecite{Ferroelectrics.326.117,Ferroelectrics.337.71}
as a high-pressure phase, is about 1~eV higher than the specific energy of the
$R{\bar 3}c$ phase, and its specific volume is higher than that of the $R{\bar 3}c$
phase (Table~\ref{table2}). This means that the large difference in enthalpies of
these phases will only increase with increasing pressure. Therefore, the $Pm3m$
phase should not be considered as a high-pressure phase. According to our
calculations, the pressure coefficient of the rhombohedral angle in the $R{\bar 3}c$
phase is $d\alpha / dP = +0.0054$$^\circ$/kbar, and so at $P = 20$~kbar the
structure remains strongly distorted, with a relative decrease of interplanar
distance of about $P/3B \approx 0.32$\%. This value is several times smaller
than the relative decrease of interplanar distance observed at the transition
pressure.

However, if one assumes that in
Refs.~\onlinecite{Ferroelectrics.326.117,Ferroelectrics.337.71} the pressure
was measured incorrectly (according to our estimates, it was underestimated by
5--7~times), and one takes the relative change of the $d_{024}$ interplanar
distance%
    \footnote{Absolute values of
    $d_{024}$ at $P = 0$ shown in Fig.~3 of Ref.~\onlinecite{Ferroelectrics.326.117}
    disagree with the lattice parameters given in this paper for the same pressure
    (the deviation is about 5\%).}
as a measure of pressure, the agreement between our calculations and experiment
becomes satisfactory. Indeed, at the $R{\bar 3}c \to Pbnm$ phase-transition
pressure (141~kbar) the calculated decrease of the $d_{024}$ interplanar distance
(compared to $P = 0$) is 2.0\%, whereas in experiment it is 2.3\%. The calculated
drop in mean interplanar distance%
    \footnote{At the $R{\bar 3}c \to Pbnm$ phase transition the (012) peak splits
    into two components with (110) and (002) indices.}
at the phase transition (0.054\%) is also close to that observed in the experiment
($\sim$0.05\%).

\begin{figure}
\centering
\includegraphics[scale=1.5]{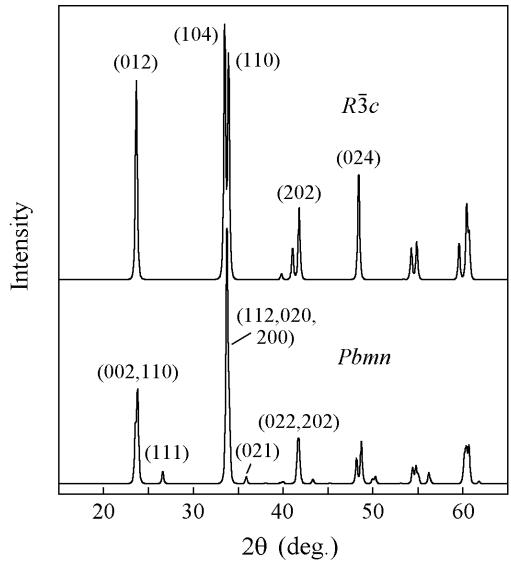}
\caption{Calculated diffraction patterns for $R{\bar 3}c$ and $Pbnm$ phases
of HgTiO$_3$ at $P = 141$~kbar (for Cu $K_{\alpha}$-radiation).}
\label{fig4}
\end{figure}

The lattice parameters and atomic coordinates for the $Pbnm$ structure at
141~kbar are given in Table~\ref{table3}. The calculated diffraction patterns
for the $R{\bar 3}c$ and $Pbnm$ phases at 141~kbar are shown in Fig.~\ref{fig4}.
The pattern for the $Pbnm$ phase is really close to that observed in the
high-pressure experiment.~\cite{Ferroelectrics.326.117,Ferroelectrics.337.71}
When transforming to the high-pressure phase, the (012) line becomes broader
because in the orthorhombic phase a pair of close lines with indices (002) and
(110) appears. The (021) line characteristic for the orthorhombic phase is
clearly seen in experimental diffraction patterns obtained during releasing the
pressure.~\cite{Ferroelectrics.326.117,Ferroelectrics.337.71}  The most serious
disagreement between our calculation and experiment consists in the absence of
(111) line of the orthorhombic phase in the diffraction patterns at high
pressure. Possibly, this is due to the incompleteness of structural transformation.
We think that new high-pressure experiments are needed to check the proposed
interpretation of the high-pressure phase transition in HgTiO$_3$.

\smallskip
\subsection{\label{sec3e}Thermodynamic stability of HgTiO$_3$}

As was mentioned in Sec.~\ref{sec3c}, the inconsistency between the observation
of SGH signal and the absence of dielectric hysteresis loops in HgTiO$_3$ at 300~K
can be explained by easiness of the defect formation. Indeed, according to
Ref.~\onlinecite{JSolidStateChem.6.509}, the samples darkened when exposed to
light. The absence of a sharp peak on the temperature dependence of dielectric
constant at the Curie temperature can also be explained by the existence of defects.
To clarify why the defect formation in HgTiO$_3$ is so easy, first-principles
calculations of the thermodynamic stability of mercury titanate were performed.

To check the thermodynamic stability of HgTiO$_3$, the enthalpy of the
$R{\bar 3}c$ phase was compared with that of the mixture of starting components,
orthorhombic HgO and rutile TiO$_2$. The calculations showed that at $P = 0$
the enthalpy of mercury titanate is 150~meV (per formula unit) higher than
the sum of enthalpies of HgO and TiO$_2$. This means that at $P = 0$ mercury
titanate is thermodynamically unstable against its decomposition into starting
components. However, because the specific volume of the HgTiO$_3$ unit cell
is significantly lower than the sum of specific volumes of HgO and TiO$_2$,
the stability of HgTiO$_3$ increases with increasing pressure. For example, at
58.8~kbar the enthalpy of HgTiO$_3$ is 75~meV lower than the sum of enthalpies
of HgO and TiO$_2$. This explains why the synthesis of mercury titanate is
possible only at high pressures.

\section{Conclusions}

First-principles calculations within the density functional theory have revealed
that the $R3c$ structure experimentally observed in HgTiO$_3$ is metastable at
ambient pressure. With increasing hydrostatic pressure, the ground-state
structure changes following the $R{\bar 3} \to R3c \to Pbnm$ sequence, and
so a new interpretation of the phase transition observed at high pressure is
proposed. It is shown that the appearance of ferroelectricity in HgTiO$_3$ at
$P = 0$ is associated with an unstable phonon mode. Optical and elastic properties
of different phases of mercury titanate are calculated. The band gap obtained
in the \emph{GW} approximation ($E_g = 2.43$~eV) agrees with the experimental data
better than the value obtained in the LDA approximation (1.49~eV). Analysis of the
thermodynamic stability explains why the synthesis of mercury titanate is
possible only at high pressures.

The calculations presented in this work have been performed on the laboratory
computer cluster (16~cores).



\providecommand{\BIBYu}{Yu}

\end{document}